\documentclass[twocolumn]{aastex701}

\usepackage[dvipsnames]{xcolor}

\newcommand{\teff}{$T_{\rm eff}$}

\begin{document}

\title{Benchmarking differential reddening in front of globular clusters}

\author[orcid=0000-0000-0000-0000,sname='Kalup']{Csilla Kalup}
\affiliation{Konkoly Observatory, HUN-REN Research Centre for Astronomy and Earth Sciences, MTA Centre of Excellence, Konkoly-Thege Mikl\'os \'ut 15-17, H-1121, Budapest, Hungary}
\affiliation{MTA–HUN-REN CSFK Lendület "Momentum" Stellar Pulsation Research Group}
\affiliation{ELTE E\"otv\"os Lor\'and University, Institute of Physics and Astronomy, 1117, P\'azm\'any P\'eter s\'et\'any 1/A, Budapest, Hungary}
\email[show]{kalup.csilla@csfk.org}  

\author[orcid=0000-0000-0000-0000,sname='Moln\'ar']{L\'aszl\'o Moln\'ar} 
\affiliation{Konkoly Observatory, HUN-REN Research Centre for Astronomy and Earth Sciences, MTA Centre of Excellence, Konkoly-Thege Mikl\'os \'ut 15-17, H-1121, Budapest, Hungary}
\affiliation{MTA–HUN-REN CSFK Lendület "Momentum" Stellar Pulsation Research Group}
\affiliation{ELTE E\"otv\"os Lor\'and University, Institute of Physics and Astronomy, 1117, P\'azm\'any P\'eter s\'et\'any 1/A, Budapest, Hungary}
\email{molnar.laszlo@csfk.org}


\begin{abstract}

Interstellar extinction is a major obstacle in determining accurate stellar parameters from photometry near the Galactic disk. It is especially true for globular clusters at low galactic latitudes, which suffer from significant amounts of, and spatially variable reddening. Although differential reddening maps are available for tens of clusters, establishing and validating the absolute zero point of relative maps is a challenge. In this study, we present a new approach to determine and evaluate absolute reddening zero-points for Galactic globular clusters by combining three-dimensional reddening maps with \textit{Gaia} DR3 RR Lyrae data. As a first case study, we investigate the low-latitude globular cluster M9. We compare the \textit{Gaia} photometry and color data of the cluster member RR Lyrae stars to field RR Lyrae stars with accurate parallaxes and whose photometric metallicities match that of M9, as well as to theoretical models. We calculate the dereddened \textit{Gaia} colors for the M9 stars based on three zero points. We confirm that the original SFD map appears to be overcorrecting the reddening for at least some RR Lyrae stars, albeit not excessively. In contrast, the 3D Bayestar map and the recalibrated version of the SFD map provide physically plausible reddenings, which we accept as lower and upper limits for M9, respectively. Our results provide a physically motivated reddening range for M9, and outline a methodology that can be directly extended to other globular clusters that are accessible to the \textit{Gaia} mission, and to other multicolor sky surveys, such as the Rubin Observatory.

\end{abstract}




\section{Introduction}
\label{sec:intro}

Globular clusters (GCs) are among the oldest constituents of the Milky Way, making them valuable fossils of our Galaxy’s formation history \citep{Massari2019}. Moreover, they not only offer opportunities for galactic archaeology but also enable detailed studies of their stellar populations, which are well known for hosting rich numbers of variable stars. One of the fundamental tools for examining GCs is the color–magnitude diagram (CMD), which can be used to separate their individual stars at different evolutionary stages, as well as to extract fundamental cluster parameters such as age, distance or metallicity. By sharing these distinctive physical properties of their members, GCs provide quasi-homogeneous samples of stars, making them powerful laboratories for studying the evolution and pulsation of low-mass stars across a wide range of evolutionary stages, from the main sequence and red giant branch to the white dwarf phase.

However, interstellar extinction can significantly complicate this picture, especially at low Galactic latitudes, which are heavily affected by interstellar dust clouds. Towards the Galactic bulge, GCs suffer not only from substantial reddening but also from variations in dust column density, producing spatially variable reddening on characteristic scales smaller than an arcminute. This so-called differential reddening across the faces of inner Galactic GCs introduces excessive broadening in their CMDs, hindering reliable determination of cluster physical parameters and the evolutionary status of their stars.

In order to deal with these issues, several de-reddening techniques have been developed to produce differential reddening maps along the fields of view of individual GCs: using photometric studies of horizontal branch stars \citep{Melbourne2004}, or calculating the average reddening for subregions of the cluster field with respect to a fiducial region \citep{Kaluzny1993, Piotto1999, vonBraun2001, Bonatto2013}. This was further improved by \cite{AlonsoGarcia2011} to a star by star basis. They assigned individual extinction values to every star in the field of view by comparing them to the ridgeline of the cluster and then used a nonparametric approximation for smoothing to remove the hard edges from the map. Another method was introduced by \cite{Milone2012}, where they defined a reference fiducial line based on the upper main sequence and/or the sub- and red giant branch to calculate the median color displacement for each star. 

Although these methods are typically used for further analysis of individual GCs, few publications aim to publish freely available differential reddening maps in a homogeneous manner, and most of them was conducted in the past two years. \cite{Legnardi2023} studied 56 Galactic GCs using multiband Hubble Space Telescope photometry. They adapted the method of \cite{Milone2012} to derive high-resolution and high-precision reddening maps for 21 targets, as well as to constrain the reddening law in their directions. Recently, \cite{Pancino2024} provided homogeneous and detailed differential reddening maps for 48 GCs based on ground-based wide-field photometry from \cite{Stetson2019}. We note that \cite{Jang2022} also investigated 43 GCs using the catalog of \cite{Stetson2019} and generated a differential-reddening catalog for 18 clusters, but we could not find these either at the stated website, or at CDS. Employing the technique of \cite{AlonsoGarcia2011}, a set of differential reddening maps were published by \cite{AlonsoGarcia2012} for the 25 brightest GCs located toward the Galactic center, a previously notoriously neglected region of the Milky Way regarding globular cluster studies. These maps are very valuable, as 15 out of the 25 clusters are not included in more recent maps, including several Messier objects such as M9, M10, M19, M28, M69 or M70. What is common in these maps is that they are \textit{relative} reddening maps, meaning that the calculated reddening values vary around zero, defined by the arbitrary reference value of the method, usually were the ridgeline of the CMD lies. This reference value needs to be determined. However, only the \cite{AlonsoGarcia2012} maps provide any absolute zero-point estimates.

Reddening maps for GC member stars have a wide range of astrophysical interest, but potentially overestimated extinction can introduce significant biases—or even lead to unphysical results. In a series of papers on inferring stellar mass loss through asteroseismology of solar-like oscillators in GCs observed by the Kepler space telescope, \cite{Maddy2025} found that using the SFD-based zero point adopted in the \cite{AlonsoGarcia2012} map resulted in unphysically low masses for early-AGB stars, and thus unrealistic ages for RGB stars—older than the Universe itself. Instead, they introduced a new approach to derive the zero point by using the latest version of the Bayestar 3D reddening map \citep{Green2019}, attempting to account only for the foreground reddening.

In this paper, we investigate the complications of accurately measuring the interstellar reddening that affects GCs, both in terms of depth and angular resolution. We discuss the limitations of global reddening maps, and how RR Lyrae-type variable stars can be employed to determine the total amount of reddening, thereby anchoring local relative reddening maps using data only from the \textit{Gaia} mission. We demonstrate this method on the globular cluster M9.

M9 (NGC 6333) is located in the constellation Ophiuchus at Galactic coordinates $b=5.54^{\circ}$ and $l=+10.71^{\circ}$, and is associated with the Gaia-Enceladus merger event \citep{Massari2025}. It is classified as a Type I cluster (showing a homogeneous heavy-element composition) with a metallicity of $\rm [Fe/H] = -1.67 \pm 0.01$ (stat) $\pm 0.19$ (sys) \citep{ArellanoFerro2013}. 
M9 is a historically neglected cluster: the most recent large-scale analysis was carried out more than a decade ago by \cite{ArellanoFerro2013}. It is well known for its rich RR Lyrae population, but also harbors Type II Cepheids and several eclipsing binaries \citep{Clement2001}. Together with M4, M80 and M19, it is one of only four GCs where solar-like oscillations have been detected and stellar masses derived for several red giants, providing a unique opportunity to constrain mass loss as a function of metallicity observationally \citep{Maddy2022,Maddy2024,Maddy2025}.

The structure of the paper is as follows. In Section \ref{sec:abs-zp}, we investigate the absolute zero point of the available reddening maps and provide an alternative procedure to infer them using three-dimensional reddening maps. In Section \ref{sec:is}, we introduce theoretical models of the instability strips of RR Lyrae stars and discuss how they can be used to constrain reddening. Utilizing only \textit{Gaia} DR3 RR Lyrae stars, Section \ref{sec:gaiaRRL} presents how we created our subsample for direct comparison with RR Lyrae stars in M9, as well as for comparison with theoretical predictions. Section \ref{sec:m9} contains our careful selection of reliable M9 RR Lyrae stars, while Section \ref{sec:res} summarizes our results. Finally, we conclude in Section \ref{sec:sum-dis}.

\section{Absolute zero-points}
\label{sec:abs-zp}

One possibility would be to use isochrones to set the cluster parameters as a reference for determining the average reddening. However, fitting an isochrone is complicated, multivariate problem, as age and chemical composition also affect the observed position of the main-sequence turnoff, whereas the shape of the red giant branch is affected by various model parameter choices, such as the mixing length and mass loss, among others \citep[see, e.g.,][]{Joyce-2018}. Since many of these parameters cannot be determined independently, the use of isochrones is quite limited in this case.

To establish the absolute extinction of these reddening zero-points, comparisons with interstellar extinction catalogs are required. \cite{AlonsoGarcia2012} provided absolute zero-point estimates for their maps by comparing them with those of \cite{Schlegel1998} (hereafter SFD). SFD is a full-sky map of the far-infrared emission from Galactic dust, combining the high-resolution IRAS measurements with the DIRBE experiment on board the COBE satellite, thereby preserving both calibration and resolution with improved accuracy. It revealed a wealth of filamentary structures on many spatial scales and at all Galactic latitudes. By converting dust temperature to dust column density, SFD serves as a tracer of Galactic extinction, normalized to the E(B--V) reddening.

However, photometry from the Sloan Digital Sky Survey (SDSS) indicated the need for a major recalibration \citep{Schlafly2010}. Using SDSS stellar spectra, \cite{Schlafly2011} confirmed that the SFD color excess should be scaled down to provide a more accurate conversion from far-infrared dust measurements to optical reddening values, E(B--V). This not only improved the accuracy by 14\%, but also favored the reddening law of \cite{Fitzpatrick1999} over that of \cite{Cardelli1989}. More recently, \cite{Chiang2023} introduced an additional correction, producing the CSFD version of the map, in which they removed contamination from extragalactic large-scale structure imprints associated with the unresolved cosmic infrared background. 

The SFD map provides the extinction integrated along the entire line of sight, whereas differential reddening maps should represent only the extinction in the foreground of the observed clusters. Maps such as SFD are therefore two-dimensional, lacking distance resolution, which can easily lead to an overestimation of the true reddening \cite[e.g.,][]{Arce1999}, even for globular clusters \citep{Hendricks2012}.

In contrast, using optical and/or near-infrared stellar photometry to trace dust makes it possible to construct three-dimensional (3D) maps of reddening. While the distance of far-infrared dust emission cannot be determined directly, stellar distances can be measured. Moreover, the relation between far-infrared optical depth and optical reddening depends on the dust composition and grain-size distribution, which vary across the Galaxy and can introduce systematic errors in the resulting maps \citep{Green2019}. The combination of \textit{Gaia} measurements of hundreds of millions of stars \citep{GaiaCollaboration2016}, probabilistic models of the dust distribution, and additional broadband photometry has produced several 3D reddening maps \citep{Green2019, Hottier2020, Leike2020, Lallement2022, Edenhofer2024, Speagle2024}. In these maps, the stellar population of the Milky Way is grouped into 3D volumetric pixels, or voxels. The angular and spatial resolutions of the maps are limited by the number of stars per voxel required to fit the models, as well as by the uncertainties in stellar distances. Since these uncertainties increase with distance, the effective resolution of the maps decreases accordingly. Furthermore, these maps have a defined reliability range: a minimum and maximum distance beyond which too few stars can be fitted—either because nearby stars are too sparse, or because distant stars are too faint and/or too few.

To correct the patchy differential reddening in front of M9, the only differential reddening map available in the literature so far is that of \cite{AlonsoGarcia2012} (hereafter AG2012). We employed three different zero points for the reddening calculation. First, we applied the zero point determined by AG2012 using the SFD 1998 map: 
$$ \rm E(B-V)_{SFD}^{1998}=0.43\,mag.$$
Then, following the suggestions by \cite{Schlafly2011}, we constructed a relationship between the recalibrated (hereafter S\&F 2011) and original (SFD 1998) reddening maps by using the coefficient of the $V$ filter in Table 6, assuming $R_V=3.1$. This resulted in a reddening of
$$ \rm E(B-V)_{S\&F}^{2011}=0.885\cdot E(B-V)_{SFD}^{1998}=0.38\,mag.$$ 
However, both of these zero points may lead to an overestimation of the measured reddening, since they include all material along the line of sight. Ideally, only the material located between the observer and the cluster should be considered. Therefore, we also revisited the technique first described in \cite{Maddy2025}, using the 3D reddening map called Bayestar 2019 \citep{Green2019}, the latest version of the Bayestar maps.

\begin{figure*}
    \centering
    \includegraphics[width=\hsize]{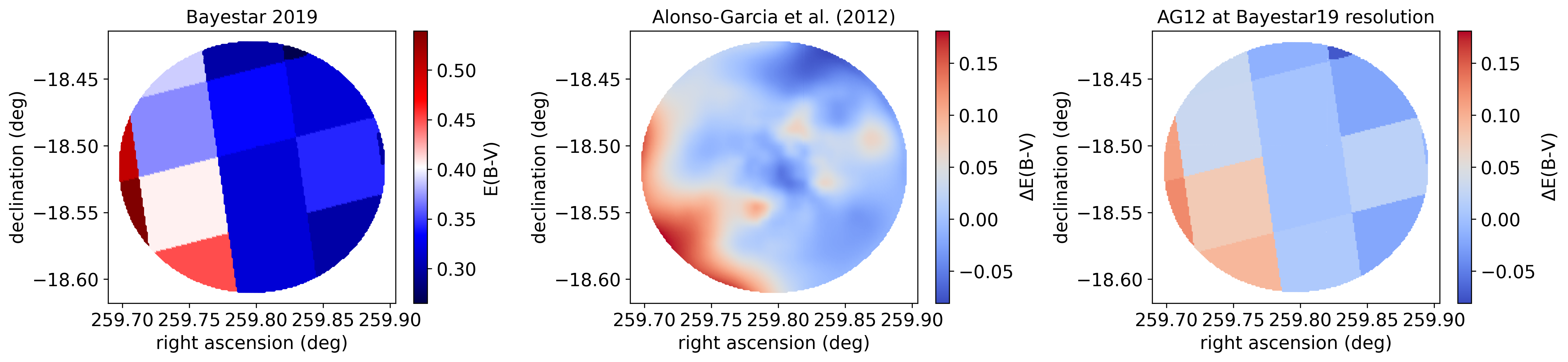}
    \caption{Differential reddening maps for M9. Left: the recreated field of view using the three-dimensional Bayestar 2019 reddening map \citep{Green2019}. Middle: the relative reddening map, as published by \cite{AlonsoGarcia2012}. Right: we degraded the resolution of the \cite{AlonsoGarcia2012} map (AG12) to the same resolution as the Bayestar pixels, as can be seen here, then compared them to set the alternative absolute zero-point for the relative map.}
    \label{fig:diffred}
\end{figure*}

We adopted the kinematic distance of $8,052\,^{+0.775}_{-0.650}\,\rm pc$ for M9 from \cite{Baumgardt2021}, which is based on \textit{Gaia} EDR3 parallaxes and thus unaffected by cluster reddening, and generated a Bayestar 2019 map for the field of view of the AG2012 map. We used the \texttt{dustmaps} Python package to assign E(g--r) reddening values to every data point in the grid of the relative map, based on their celestial coordinates and the M9 distance, using the 'median' query mode to return the median reddening (although we found no significant difference with the other options). Figure \ref{fig:diffred} shows the resulting comparison, also highlighting the resolution difference between the two maps, and thus the importance of the relative maps. Following the method described by \cite{AlonsoGarcia2011,AlonsoGarcia2012}, we degraded the resolution of the relative map to match the coarser Bayestar resolution, averaging to obtain one extinction value per Bayestar pixel (right panel of Figure \ref{fig:diffred}). For this, we also queried the HealPix information of the Bayestar map, enabling us to divide the AG2012 stars into the same pixels as the Bayestar grid displays. Before the comparison, we converted the Bayestar reddening via $\rm E(B-V)=0.883\, E(g-r)$, following Equation 29 of \cite{Green2019} and the recommended conversion of $\rm E(B-V)=0.98\cdot E(g-r)$ in \cite{Schlafly2011}. We then fitted a linear equation with a unit slope to determine the Bayestar 2019 absolute reddening at the zero-point of the AG12 map. Two pixels were excluded from the fit due to their poor coverage in the AG12 map. The Bayestar values for these pixels showed a large deviation because they were dominated by stars outside the field of view, which would have introduced a bias. The resulting offset between the two maps at the same resolution was
$$ \rm E(B-V)_{bayestar}^{2019}=0.33\,mag,$$ 
significantly lower than the previously reported (0.43) or calculated (0.38) values from SFD. However, the issue with this Bayestar 2019 estimate is that the distance of M9 lies just beyond the reliability range of the 3D map along this line of sight. Although the difference between the cluster distance and the map’s maximum distance is small, it raises the possibility of at least a slight \textit{underestimation} of the true reddening. In order to evaluate the absolute zero-point estimates, we applied them to the RR Lyrae stars of M9 and checked whether they fall into the expected positions within the predicted instability strip, or if there is a clear contradiction with either of them.

\section{Instability strips of RR Lyrae stars}
\label{sec:is}

RR Lyrae (hereafter RRL) stars populate the intersection of the horizontal branch and the classical instability strip, marking a key zone on the Hertzsprung-Russell diagram (HRD). RRLs can be divided into three groups based on their light curves and pulsation modes: RRab stars pulsate in the radial fundamental mode, RRc stars pulsate in the first overtone. These two groups make up most of the class: the remaining small fraction is the RRd stars, which exhibit both modes simultaneously \citep{Clementini2023}.

RRL stars can be used to determine the distances to globular clusters as well as the interstellar extinction along the line of sight. However, these two quantities have not always been disentangled from each other, and their combination is referred to as the apparent distance modulus, defined as the difference between the apparent and absolute brightness of an object ($\mu = m-M$). To determine the true distance modulus ($\mu_0 = m-M-A$) and separate the effects of interstellar extinction, we need to identify observable properties against which interstellar absorption and reddening can be calibrated.

One of the oldest methods is Sturch's method, or the minimum-light method, which is based on the observation that RRab stars have (almost) the same colors at minimum light, which can be then used to determine the reddening in the measurements. \citet{Sturch-1966} showed this for Johnson (\textit{U--B}) and (\textit{B--V}) colors. This technique was later extended to (\textit{V--R}) and (\textit{V--I}) colors and period and metallicity terms were added \citep{Walker-1990,Blanco-1992,Guldenschuh-2005,Kunder-2010}. Both observational and theoretical works showed that the period-color relation of RR Lyrae stars is flat at minimum light, supporting these applications \citep{Bhardwaj-2014,Das-2020}. This work was then extended even further by \citet{Piersimoni-2002} who also included the pulsation amplitudes and derived an amplitude-color-metallicity relation for the cluster stars. 

A drawback to Sturch's method and its derivatives is that they rely on time-resolved multicolor observations to determine minimum brightness of the stars in specific passbands. Here we present an approach based on CMDs and color information, which directly utilizes the \textit{Gaia} data of field and cluster stars to derive the dereddened CMD at a selected metallicity range for well-characterized field RRL stars. The method then scales the star of the cluster directly to this CMD to determine the allowed range of interstellar extinction values in front of the cluster.

Our analysis depends on the shape and extent of the RRL instability strip in the HRD. Along the temperature axis it is limited by the highest  \teff{} at which the partial ionization zone is deep enough in the envelope to drive pulsations, and by the lowest \teff{} before convective motions become too vigorous to maintain coherent pulsation modes in the envelope. Along the luminosity axis, it is limited by the position of the ZAHB and the end of core He burning at the bottom and top, respectively. Of course, the exact positions of these boundaries limits depend on many physical parameters, including the chemical compositions of the stars.

The structure of the instability strip (hereafter IS) can be mapped both by pulsation models or by observations. While linear models can be computed fast, they only map the locations of possible excited modes via the mode growth rates. Non-linear models are time-consuming to compute but they can tell us which modes develop into large-amplitude pulsation. Various authors mapped the blue and red edges of the first-overtone (FOBE and FORE) and fundamental (FBE and FRE) modes \citep[see, e.g.,][]{Bono-1995,Bono-1997,Marconi-2003,DiCriscienzo-2004,Marconi2015}. While the positions of the outer edges (the FOBE and FRE) depend on various modeling choices such as chemical composition (elemental abundances and opacities) and convective parameters, their positions are relatively well established. The inner edges (the FORE and FBE), however, are also affected by mode selection mechanisms. The transition between fundamental-mode and overtone pulsation includes both a double-mode region and an either-or region where single-mode pulsation in either mode is possible. \citet{Bono-1995} hypothesized that this hysteresis region is connected to the Oosterhoff dichotomy. Later, \citet{Szabo-2004} mapped these regions, concluding that the hysteresis depends on the direction of evolution. However, the validity of double-mode pulsations in 1D pulsation models remains debated \citep{Smolec-2008}.

The IS can also be mapped observationally, offering a way to test the model calculations, but that requires accurate distance and extinction measurements for every star. Historically, this was circumvented by mapping the IS of GCs, especially the large RRL populations of M3 and M5 \citep[see, e.g.,][]{Caputo-1999,Cacciari-2005,Kumar-2024}. For field stars, however, the breakthrough came with the \textit{Gaia} mission \citep{GaiaCollaboration2016}. With accurate parallaxes available for thousands of RRL stars, in conjunction with detailed extinction maps, it became possible to map the IS with field stars, as well \citep{Clementini2023}, including the separation of RRc and RRab stars, with the double-mode stars falling into the middle \citep{Molnar-2022}. Large-scale spectroscopic surveys can also be used for such purposes \citep{Medina-2025}. Recently, \citet{Cruz_Reyes-2024} mapped the population of GC member RR~Lyrae stars in 75 clusters using \textit{Gaia} DR3 data.

The most recent in-depth modeling analysis of the RRL IS was conducted by \citet{Marconi2015}. They determined the edges for a range of Z metallicity values. For each Z, three model sequences were computed in 100\,K steps. The initial sequence was based on the $\alpha$--enhanced ZAHB models calculated by \citet{Pietrinferni-2006}, with the mass selected at the mid-IS point (at $\log{}T_{\rm teff} =3.85$). The two other sequences were calculated at luminosities $+0.1$\,dex above the ZAHB and at the end of core He burning. Unlike the evolutionary models, the pulsation models then used an older solar mixture, with $Z_\odot = 0.02$ \citep{Grevesse-1993}. We chose the $Z=0.0003$ model sequences as those are the closest to the [Fe/H] index of M9, when using the modern solar value of $Z_\odot=0.014$. 

In order to convert the \citet{Marconi2015} edges into the \textit{Gaia} CMD plane, we first transformed the \teff{} values to $(BP-RP)$ colors using the inverse of the color conversion published by \citet{Mucciarelli-2021}. We then transformed the luminosities into $M_{\rm G_0}$ magnitudes using the \texttt{gaiadr3\_bcg} tool, which calculates the bolometric correction in the \textit{G} band for a given \teff{}, [Fe/H], [$\alpha$/Fe] and $\log{}g$ value parameter set \citep{Creevey-2023}. We assumed $\log{}g = 3.0$ and 2.3 for the FOBE and FORE, and $\log{}g=2.4$ and 1.5 for the FBE and FRE, respectively. The $\log{}g$ values were chosen based on the collection of high-quality spectroscopic data for RR Lyrae stars in \citet{Molnar-2023,Molnar-2024}, at the average \teff{} values of the various edges. For comparison purposes, we also converted the FOBE and FRE edges calculated by \citet{Marconi-2003}. These are slightly redder than the later edges, but their uncertainty ranges overlap.

More recently, \citet{Cruz_Reyes-2024} calculated IS edges based on a grid of linear MESA-RSP models. The RSP models are not tied to evolutionary tracks, thus they could extend the grid to luminosities lower than those used by \citet{Marconi2015}. However, linear models only provide mode growth rates. This leads to uncertainties whether the models develop into true pulsations, and into which mode if both modes are linearly unstable. They also varied the He content of the models, and found some discrepancies between the RRc and RRab samples, with the highest He content fitting the former and the lowest He content fitting the latter group the best. Given these ambiguities, we decided not to use their IS boundaries. 

\section{\textit{Gaia} DR3 RR Lyrae sample}
\label{sec:gaiaRRL}

We used the Cepheid and RR Lyrae sample in the \textit{Gaia} DR3 catalogue produced by the Specific Object Study pipeline (hereafter SOS Cep\&RRL or SOS RRL sample). Currently, the SOS Cep\&RRL sample is the largest and most homogeneous all-sky catalog of RR~Lyrae and Cepheid variable stars in the brightness range 7.64--21.14~mag \citep{Clementini2023, Ripepi2023}. The SOS Cep\&RRL sample contains a cleaned and validated set of 270,905 RR Lyrae stars across the whole sky, including 95 globular clusters and 25 Milky Way companions (such as the Magellanic Clouds and several dwarf galaxies). The number of confirmed RRL stars nearly doubles that of the DR2 RR~Lyrae catalogue \citep{Molnar2018, Rimoldini2019, Clementini2019}. This sample was released with multiband ($G$, $G_{\rm BP}$, $G_{\rm RP}$) epoch photometry (light curves), 1,096 of which are also supported by radial velocity time-series measurements.

The SOS Cep\&RRL catalog is accessible via the \textit{Gaia} archive\footnote{\url{https://gea.esac.esa.int/archive/}} using the \texttt{gaiadr3.vari\_rrlyrae} table, which contains characteristic parameters of the light curves derived by the SOS pipeline, such as peak-to-peak amplitudes of the $G$, $G_{\rm BP}$, and $G_{\rm RP}$ light curves, pulsation periods, and mean magnitudes 
along with the $\varphi_{21}$, $\varphi_{31}$, $R_{21}$, and $R_{31}$ parameters from the Fourier decomposition of the $G$-band light curves. The RRL sample was also classified by type/pulsation mode (RRab, RRc, or RRd): 174,947 fundamental-mode, 93,952 first-overtone, and 2,006 double-mode RRL are included in the catalog. Photometric metallicities ([Fe/H]) were derived from these Fourier parameters by applying a relation between the pulsation period and the $\varphi_{31}$ Fourier parameter \citep{JK-1996, Nemec-2013}, and were released for 133,559 RRL in the catalog. However, the calibration of \citet{Nemec-2013}, used in DR3, was found to be inaccurate for long-period and/or low-amplitude RRab stars. Therefore, we used the reprocessed $\rm [Fe/H]$ metallicity values from a more recent catalog by \cite{Muraveva-2025}, who presented a new, machine learning-based method to calculate individual metallicities based on the periods and Fourier parameters of the aforementioned \textit{Gaia} $G$-band light curves, combined with spectroscopic metallicity measurements from the literature.

We defined our initial sample as all the RRL stars with valid metallicity estimates in the Muraveva \& SOS joint catalog, which included 114,768 RRab and 20,001 RRc stars. To compare them with stars from M9, we filtered this sample to include only stars within the metallicity range of M9. As only a few metallicity measurements exist in the literature for M9, and these values vary, we chose a conservative metallicity range following \cite{ArellanoFerro2013}. Stars with $\rm [Fe/H] = -1.67\pm0.19$ resulted in 40,215 RRab and 5,353 RRc stars. We then queried additional parameters from the \textit{Gaia} archive for these stars, such as brightnesses from the \texttt{gaiadr3.gaia\_source} table, the BP/RP excess factor \citep{Riello2021}, and various distances from the joint table \texttt{external.gaiaedr3\_distance}. This table represents the latest version of the catalogs of \cite{Bailer-Jones2021}, containing several types of distance estimates. We used the geometric distance, which relies solely on \textit{Gaia} parallaxes and uncertainties, along with a Galactic model prior. To place these stars in an absolute frame, we needed to determine their true absolute $G$ magnitudes and true \textit{Gaia} $(BP-RP)$ colors, which requires proper stellar extinction and reddening estimates.

As discussed above, it is crucial to consider only foreground extinction to avoid overestimating reddening. This is especially important for field stars in this catalog, which span a much larger range of distances compared to typically more distant GCs. First, we applied a quality cut on the \texttt{parallax\_over\_error} parameter at $\varpi/\sigma_\varpi > 10$, to select the best candidates for our final sample \citep[see also][]{Vasiliev2021}. This choice balances the need to retain high-quality data with the desire not to lose too many stars, resulting in 814 RRab and 163 RRc stars. 
As before, we retrieved reddening estimates from the Bayestar 2019 catalog based on spatial coordinates and excluded all stars outside the map’s reliability range. To minimize the impact of incorrect reddening corrections, we initially selected stars with low reddening, applying an additional filter of E(B--V) $<$ 0.1, but for the same reason we also excluded stars with exact zero values. After this step, our final sample consisted of 814 RRab and 49 RRc stars.

We calculated the absolute $G$ magnitude as follows:
$$ M_{\rm G_0} = G_0 + 5 - 5\, \log_{10}(r),$$
where $G_0 = G-A_G$ is the extinction-corrected $G$ magnitude, and $r$ is the distance. For all extinction-corrected \textit{Gaia} magnitudes ($G_0$, $G_{\rm BP_0}$ and $G_{\rm RP_0}$) we used the official \textit{Gaia} DR3 extinction law\footnote{https://www.cosmos.esa.int/web/gaia/edr3-extinction-law} for giant stars, following the recommended procedure. Assuming $A_0 = R_VE(B-V)$ with $R_V=3.1$, after a few iterations the $k$ extinction coefficients clearly converge. Finally, we derived the true colors as:
$$ (BP-RP)_0=G_{\rm BP_0}-G_{\rm RP_0} $$

\begin{figure}
    \centering
    \includegraphics[width=\hsize]{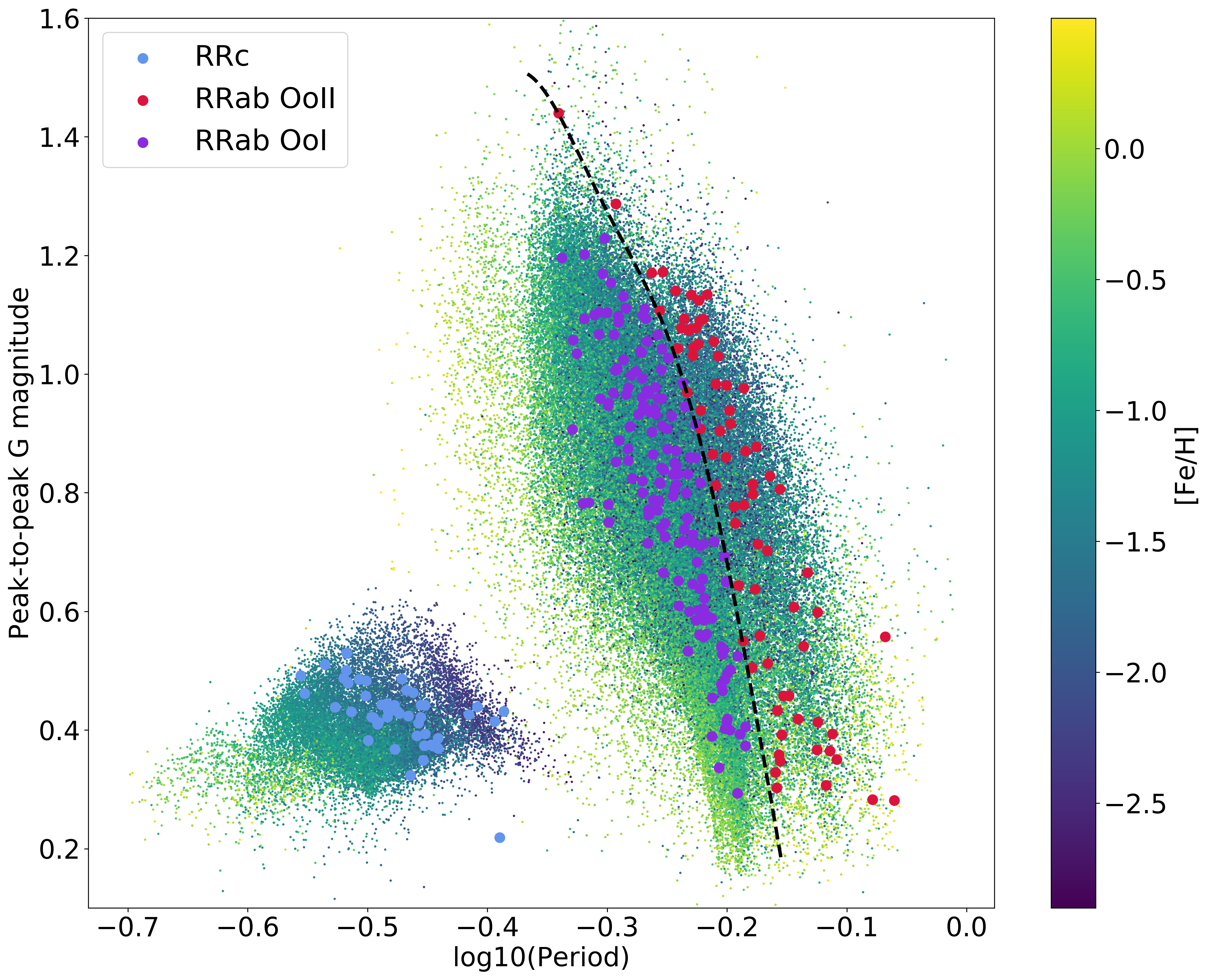}
    \caption{Bailey diagram for the \textit{Gaia} DR3 SOS RR Lyrae catalog \citep{Clementini2023}. It uses the recalculated photometric metallicities from \cite{Muraveva-2025} and also shows our filtered RRab (purple and red) and RRc (light blue) sample stars. The Oosterhoff separation is indicated by a dashed black line based on \cite{Luongo-2024}. Purple dots mark Oosterhoff I-type RRab stars, while red dots correspond to Oosterhoff II.}
    \label{fig:Oo}
\end{figure}

Because M9 is an Oosterhoff II (OoII) type cluster, we also divided our SOS RRab sample into OoI and OoII groups. We used the relation derived by \cite{Luongo-2024} for \textit{Gaia} periods and amplitudes to separate our RRab stars in the period–amplitude plane. Our selected stars from the SOS \& Muraveva joint RRL sample, showing also the RRab division, are presented in Figure \ref{fig:Oo} in their Bailey diagram.

\section{M9 RR Lyrae sample}
\label{sec:m9}

First, we collected the known RRab and RRc stars for M9 from the Catalogue of Variable Stars in Globular Clusters (CVSGC) by \cite{Clement2001}. However, as \cite{Prudil2024} pointed out, not all previously identified M9 RRL stars are actual cluster members. Using \textit{Gaia} EDR3 kinematic properties of globular clusters, cluster member selection of nearby stars was performed by \citet{Vasiliev2021} to assign membership probabilities. \cite{Prudil2024} crossmatched stars in the CVSGC with the analyzed sample of \cite{Vasiliev2021} and found that globular clusters located toward the Galactic bulge are particularly affected by field stars, which is unsurprising given the large number of fore- and background populations in the Bulge region. This is exactly the case for M9 as well. From the listed 21 RRL stars, only 11 are actual cluster members: 4 RRab (V1, V2, V4, V7) and 7 RRc (V5, V9, V10, V18, V19, V22, V23). With the exception of V19, which received 70\%, all have a probability rating of 90\% or 100\%. We used these 11 stars for further investigation in our analyses. We matched the \textit{Gaia} DR3 \texttt{source\_id} with each of the member stars and queried their \texttt{gaiadr3.gaia\_source} catalog mean brightnesses and quality flags (most importantly \texttt{phot\_bp\_rp\_excess\_factor}). We also crossmatched this sample with the SOS RRL catalog and found 10 out of 11 stars, with only V5 missing. 

It is not surprising that there is a difference between the main \textit{Gaia} DR3 and the SOS RRL catalog in the reported brightnesses. In the DR3 catalog, brightness is computed as the weighted mean flux of the given passband, without accounting for the pulsation cycle. In contrast, the SOS catalog provides a true intensity-averaged magnitude. Another method to derive mean magnitudes from light curves of RRL stars is based on the zero point of the Fourier decomposition analysis, which is also available for some of our stars in the SOS RRL catalog. To validate these estimates, we downloaded the \textit{Gaia} DR3 epoch photometry of our sample stars directly from the \textit{Gaia} archive and performed our own Fourier fits (in magnitude space) for the 10 stars. We note that, especially for large-amplitude pulsators, intensity-averaged and magnitude-averaged means are not equal. The difference becomes more prominent in RRab stars due to their asymmetric light curves and larger amplitudes, reaching the highest values in the $G_{\rm BP}$ band. This effect was already reported in several previous works \citep[e.g.][]{Marconi2006}. Therefore, we adopted the intensity-averaged mean magnitudes, which minimize the difference.

\begin{figure*}
    \centering
    \includegraphics[width=\hsize]{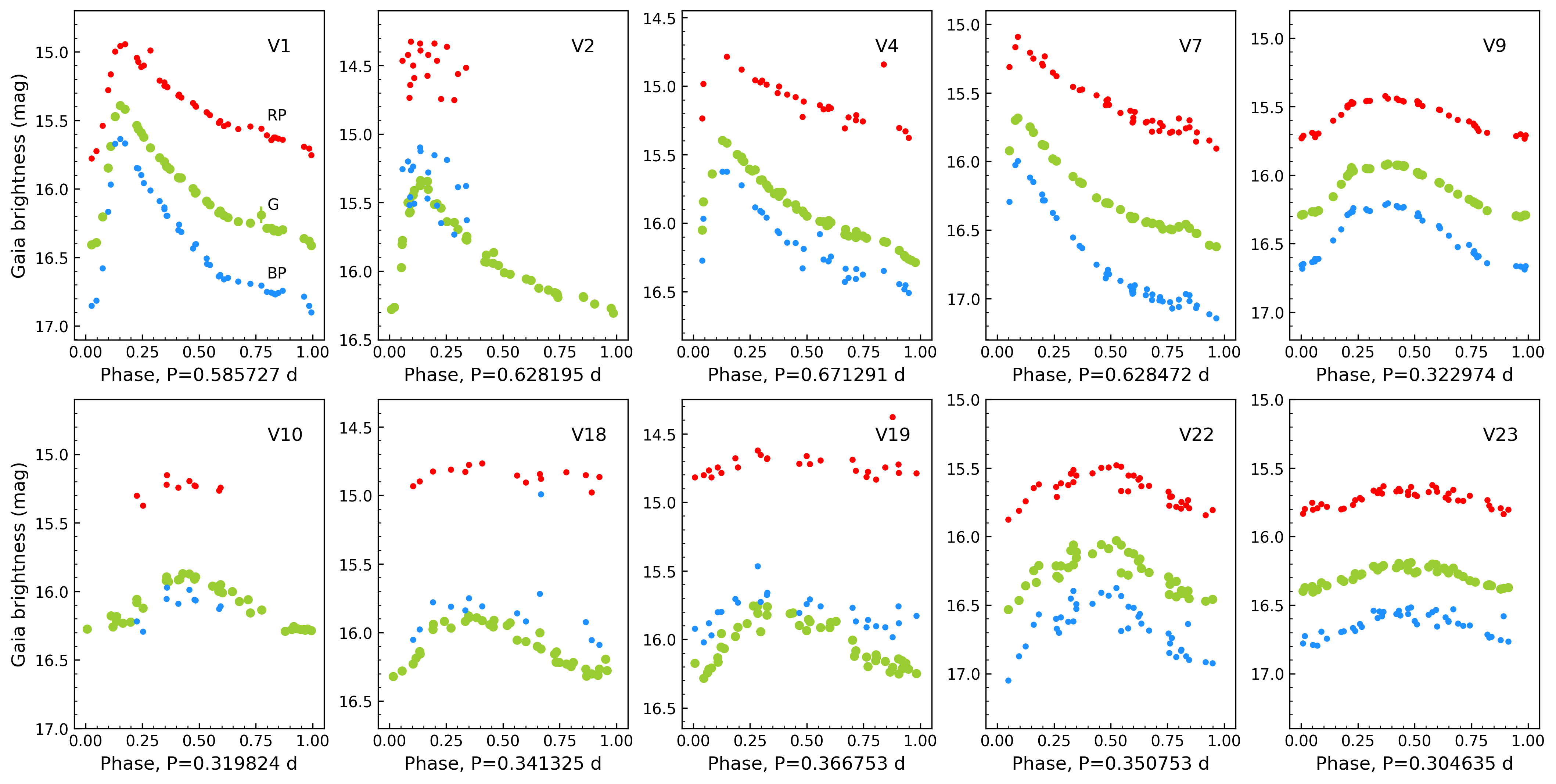}
    \caption{Phase folded light curves of the M9 RRL sample, based on \textit{Gaia} epoch photometry and variability periods. Red, green and blue points correspond to $G_{\rm RP}$, $G$ and $G_{\rm BP}$ passbands, respectively. Note that for V2 and V10, the phase coverage of the red and blue bands are very poor, while for V2, V10, V18 and V19, the $G_{\rm BP}$ data are very close to, or above the $G$ band points. In both cases, they lead to incorrect estimates of the mean magnitude.}
    \label{fig:lcs}
\end{figure*}

In Figure \ref{fig:lcs}, we present all the available \textit{Gaia} light curves for our M9 sample. We compared the \textit{Gaia} periods with those from the literature and found no significant differences, so we adopted the \textit{Gaia} values to construct the phase-folded light curves. There are two crucial aspects to consider regarding these light curves. First, they need to be well covered in phase; otherwise, their averages will not reflect the true shape of the light curve. This is the case for V10, for which the SOS catalog could not provide any $G_{\rm RP}$ and $G_{\rm BP}$ estimates. A similar situation occurs for V2, except where an estimate is available but clearly incorrect, as it samples only the peak magnitudes of its $G_{\rm BP}$ and $G_{\rm RP}$ light curves. Second, based on the general properties of RRL stars, we expect that at shorter optical wavelengths the stars appear fainter, meaning that the $G_{\rm BP}$ magnitudes should lie \textit{below} the $G$ magnitudes. For V2, V18, and V19, a clear discrepancy is visible in the \textit{Gaia} measurements, while for V10 the $G$ and $G_{\rm BP}$ magnitudes appear suspiciously close to each other.

Luckily, there is a way to test whether the $G_{\rm BP}$ and $G_{\rm RP}$ photometry is consistent with the $G$ photometry. \cite{Evans2018} defined an indicator called the BP and RP flux excess factor ($C$, queried as \texttt{phot\_bp\_rp\_excess}), based on the ratio between the total flux in BP and RP and the $G$-band flux. However, this quality metric shows a strong dependence on color, so \cite{Riello2021} introduced the corrected color excess factor $C^*$, for which ideally $C^* \approx 0$. We computed $C^*$ for all stars in our M9 RRL sample to quantify possible inconsistencies and found that V1, V7, V9, V22, and V23 lie below $C^* < 0.003$, within the $3\sigma$ scatter for a sample of well-behaved isolated stellar sources with good-quality \textit{Gaia} photometry \citep[see Eq. (18) in][]{Riello2021}. In contrast, the other M9 stars in our sample show much higher values ($0.2\leq C^* \leq 1.4$). We therefore considered them problematic and restricted our subsequent analysis to the reliable sources V1, V7, V9, V22, and V23. We list all the aforementioned parameters for each star in Table~\ref{table}.

For the reddening correction, we employed the AG2012 relative map. Based on right ascension and declination, we identified the closest grid point for each of our M9 stars and assigned an E(B--V) value in the same way as described in Section \ref{sec:abs-zp}. We applied the original and recalibrated SFD, as well as the Bayestar 2019 absolute zero points to them. We then converted these values into the \textit{Gaia} passbands using the method presented in Section \ref{sec:gaiaRRL}, and used them to calculate $M_{\rm G_0}$ and $(BP-RP)_0$ for both reddening solutions.

\section{Results}
\label{sec:res}

\begin{figure*}
    \centering
    \includegraphics[width=\hsize]{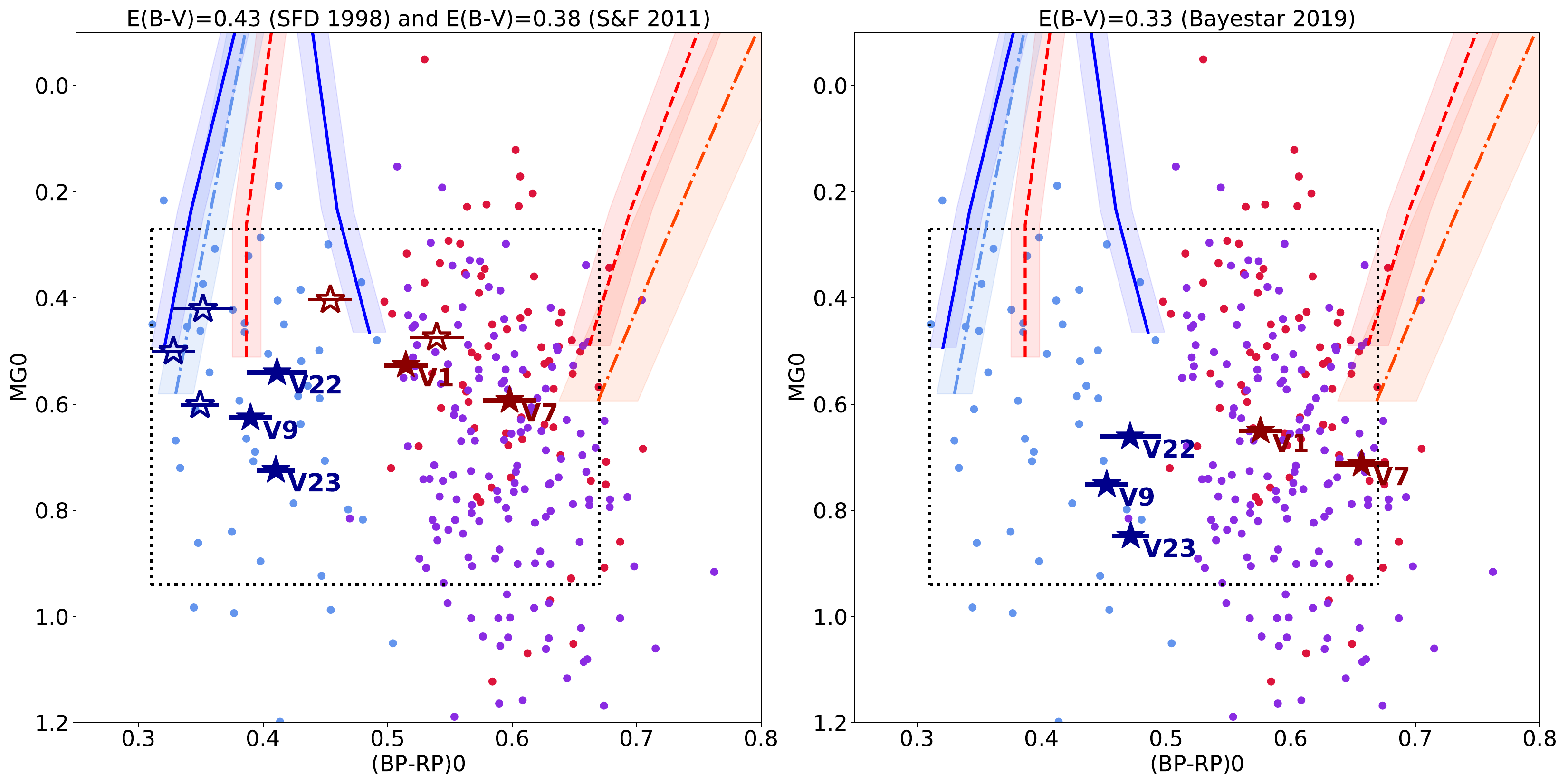}
    \caption{The RRL CMD at the metallicity of M9. Small points represent the \textit{Gaia} sample, with blue, purple and red corresponding to RRc, OoI RRab and OoII RRab stars, respectively. The origins of low-luminosity points are discussed in the text, and the dotted square corresponds to the observed RRL distribution of GCs \citep{Cruz_Reyes-2024}. The solid and dashed lines indicate the \cite{Marconi2015} instability strip (IS) edges, for RRc (blue) and RRab (red) models. The dashed-dotted light blue and orange lines represent the outer edges of the RRL IS published by \cite{Marconi-2003} and \cite{DiCriscienzo-2004}. Large star symbols are the M9 RR Lyrae stars for which we found the \textit{Gaia} color data to be accurate. In the left and right panels we show these M9 stars using the SFD 1998 (empty), S\&F 2011 (filled) and Bayestar 2019 reddening zero points, respectively. The Bayestar and the S\&F 2011 values places the RR Lyrae stars within physically plausible limits, serving as reasonable bounds for the reddening in front of M9, whereas the SFD 1998 zero point shifts one of the RRab stars into the RRc instability strip, supporting the idea that it causes an overestimation.}
    \label{fig:IS}
\end{figure*}

In Figure \ref{fig:IS}, we present a comparison of our final M9 sample (large star symbols) and our filtered SOS RRL sample (light blue, purple, and red dots) with the theoretical instability strips by \cite{Marconi2015} (blue for the RRc region, red for the RRab region), as well as by \cite{Marconi-2003} and \cite{DiCriscienzo-2004} (dashed-dotted lines). The left panel shows a reddening correction based on the SFD 1998 (empty) and S\&F 2011 (filled) reddenings, while in the right panel we applied the absolute zero point inferred from the 3D Bayestar 2019 reddening map.
    
\subsection{Comparison with \textit{Gaia} field stars and the theoretical instability strips}

When we compare the dereddened $(BP-RP)_0$ colors to the Marconi instability strip edges, we find that the RRc edges fit the observations very closely. The division between RRc and RRab stars follow the FORE: we find virtually no RRab stars blueward of it, despite the bluer position of the model FBE. RRab stars also agree with the edges quite well, although the observed red edge extends slightly beyond the FRE, especially at low luminosities. Since the red edge is defined by the properties of the convective zone of the models, this small discrepancy could come from convective parameter choices, and also from underestimated reddening corrections due to underestimated distances, as discussed below. These differences, however, are small, and will not hinder our analysis.

In this CMD, the \textit{Gaia} points extend below the brightness of typical ZAHB models: for example, the chemical composition of M9, the BaSTI (Bag of Stellar Isochrones, \citealt{Hidalgo-2018,Pietriferni-2021}) model library predicts ${\rm G_{ZAHB}}\approx0.5$\,mag. The same effect is visible in the RR Lyrae CMDs calculated by \citet{Molnar-2022}, as well. The origins of this discrepancy are likely be multi-faceted. From the theoretical side, certain model considerations, such as convective parameters, elemental opacity values, atomic diffusion, mass loss all can influence the calculated ZAHB luminosity, as it was shown by \citet{Pietriferni-2021}, for example. From the observational side, uncertainties in interstellar extinction can affect the absolute brightness values. We note that \citet{Cruz_Reyes-2024} found that while most GC stars are brighter than 0.7 mag in $M_{\rm G}$, the 5th to 95th percentile range extends to $0.94 \lesssim  M_{\rm G}<0.31$ mag, which extends below the ZAHB models. We find even fainter targets among the field stars: here a likely additional effect is coming from the method used by \citet{Bailer-Jones2021} to calculate the distances. The Bayesian prior employed in their calculations moves stars toward the densest part of the Galactic prior, proportionally to the relative uncertainties in the parallax measurements. This implies that distances of nearby stars (closer than the peak of the density distribution) with uncertain parallaxes may be overestimated, while distances to more distant stars may be underestimated. We do observe an increase in less luminous stars beyond $\sim2.3$ kpc, which suggests that the distances of more distant RR Lyrae stars may indeed be slightly underestimated in the catalog. 

Because of these, we elected not to compare the M9 and \textit{Gaia} stars with ZAHB models, and instead relied only on the distribution of the stars in color. Colors are affected by distance only through the amount of reddening they suffer, and small systematic errors in distance translate either into very small shifts in reddening, or possibly none at all if the stars remain in the same map voxel, depending on the decreasing line-of-sight resolution with increasing distance in 3D maps. 
    
We also note that we do not see any difference between the OoI- and OoII-type RRab stars regarding their positions within the instability strip.

\subsection{Comparison of the M9 and \textit{Gaia} samples}

We placed the reliable M9 stars (V1, V7, V9, V22, and V23) onto the CMD in Figure \ref{fig:IS} with all of the zero-point options. The Bayestar 2019 solution places the M9 RRc stars to the FORE, but still certainly within the RRc locus, while the RRab stars fall well inside the expected fundamental-mode region. Applying less reddening would push the RRc stars into the RRab regime. Therefore, we confirm that value inferred from the 3D Bayestar map is physically plausible, and we consider it as a lower limit on the absolute reddening of M9.

The SFD 1998 zero point shifts the RRc stars very close to, but still within, the FOBE. However, one of our RRab stars (V1) has now been moved out of the RRab region and falls among the RRc stars instead. While it remains redward of the calculated FBE, the distribution of field stars suggest that RRab stars do not extend beyond the FORE. The position of V1 indicates that the SFD 1998 zero point overestimates the absolute reddening of the cluster, albeit not excessively. This result agrees with the findings of \citet{Maddy2025}, who found that this reddening and the associated distance modulus in the literature leads to unphysically low asteroseismic masses for the cluster red giants.

\section{Discussion and summary}
\label{sec:sum-dis}
    
In this paper, we present a new approach to infer and/or evaluate absolute reddening zero points for Galactic globular clusters using three-dimensional reddening maps and/or \textit{Gaia} DR3 RR Lyrae stars. We investigate the case of the low-latitude, inner globular cluster M9, which is affected by both significant amounts of interstellar extinction and by significant spatial variations across its face. Although a single differential reddening map is available for this cluster in the literature, the only provided absolute zero point for that extinction map is based on a comparison with the two-dimentional SFD far-infrared reddening map from 1998 \citep{AlonsoGarcia2012}. That paper warns that their SFD-based estimates include extinction not only in the foreground but also in the background of the observed clusters, and furthermore, \cite{Maddy2025} also reported physical contradictions when applying the provided SFD 1998 zero point to M9.

Here, we revisit the use of three-dimensional reddening maps, particularly the Bayestar 2019 map \citep{Green2019}, to demonstrate an alternative method for calculating zero points for any relative differential reddening map in the literature. We also introduce a technique that compares RR Lyrae stars from a given globular cluster to a filtered sample from the \textit{Gaia} DR3 SOS catalog to derive absolute reddening constraints. The method does not require new measurements to be collected, avoids the limitations of global dust maps, and can be applied to any cluster containing an RR Lyrae population accessible to \textit{Gaia}.

The demonstration on M9 clearly shows the potential to constrain physically reliable interstellar reddenings in front of globular clusters. We tested the derived Bayestar and SFD absolute zero points for M9 in this way and found that using the Bayestar value places the RRab and RRc stars within their expected pulsational regimes, confirming it as a plausible solution. Because the M9 RRc stars fall close to the red edge of the first-overtone instability strip, this can also be considered as the lower limit for the reddening in front of the cluster. In the case of the SFD dust map, we used not only the original 1998 value, but also its inferred recalibrated 2011 version. When we employed the SFD 1998 zero point, the RRc stars fall very close to the blue edge of their instability strip, and one of the M9 RRab stars is shifted into the RRc regime instead of the RRab one, concluding that it clearly overestimates the total reddening M9 experiences. Therefore, our limits for the true value are $0.33 \le {\rm E(B-V)} \le 0.38$\,mag, constrained by the Bayestar 2019 and S\&F 2011 estimates, respectively.
We note that the comparison with the field RR Lyrae stars and the theoretical models chosen for the same metallicity range, as well as with the distribution of the observed GC RRLs in the color-magnitude diagram shows good agreement.

Although 3D dust maps offer a unique opportunity to estimate reddening more accurately, they are also limited by the extinction law. Even though it may seem that color excess is used directly, reddening maps were constructed based on \textit{underlying assumptions} about $R_V$ to derive reddening, usually treating the reddening law as constant throughout the Galaxy. The Bayestar 2019 map assumes a universal dust extinction law with $R_V\approx 3.1$ and the authors emphasize the need for future efforts to allow the dust extinction spectrum to vary in order to construct more accurate 3D maps \citep{Green2019}. Currently, it is not possible to manually change the map to an arbitrary $R_V$ value, but the next generation of reddening maps may be just around the corner. Recently, \cite{Zhang2025} mapped the three-dimensional variation of the extinction curves, providing $R(V)$ parameters for the foreground dust as a function of distance along any given line of sight in the Milky Way.
    
We also draw attention to the possible inaccuracies in the published \textit{Gaia} and SOS mean brightness values. The flux-averaged magnitudes in the main catalog are not always accurate for large-amplitude pulsators, as taking the pulsation cycle into account is unavoidable. For stars flagged as variables, the SOS catalog provides two types of mean magnitudes: intensity- and magnitude-averaged values (the latter based on Fourier decomposition). We note that these are not equal, and the magnitude-averaged values can cause a systematic offset, which becomes more prominent for RRab stars; hence, we also recommend using the intensity-averaged magnitudes. For a sanity check, downloading the epoch photometry and checking the phase coverage is also recommended, because in some cases we found that the published brightness values correspond to false averages due to incomplete cycles. Additionally, calculating the corrected color excess defined by \cite{Riello2021} for the variables ensures that there is no inconsistency between the measured $G$-band, BP and RP fluxes.

Our results can be used not only for further direct analyses related to M9, but can also be adapted to other Galactic globular clusters. Furthermore, the method can also be adapted to future wide-sky survey missions collecting multicolor photometry of globular cluster stars, such as to data coming from the Rubin Observatory.


\begin{deluxetable*}{lcccccc} 
\tablewidth{0pt} 
\tablecaption{Brightnesses and other arameters of the five reliable RR Lyrae stars. \label{table}} 
\tablehead{ 
 \colhead{Variable ID} & \colhead{~} &  \colhead{V1}  &  \colhead{V7}  &  \colhead{V9}  & \colhead{V22}  & \colhead{V23}}
\startdata
Type  & & RRab  & RRab  & RRc   & RRc   & RRc   \\
\textit{Gaia} DR3 source\_id  & &4122470693492988032  &4134456778975963904  &4122469422182483968  &4122446641654250496  &4122470895318178048  \\
 RA  &(deg) &259.82629 &259.767882 &259.897479 &259.762954 &259.838085 \\
 Dec  &(deg) &-18.53935 &-18.54013 &-18.57145 &-18.59201 &-18.50287 \\
 Period  &(d) &0.585727 &0.628472 &0.322974 &0.350753 &0.304635 \\
E(B--V)  &(mag) &0.326 &0.398 &0.325 &0.443 &0.367 \\
 E(B--V)$_{\rm SFD}$  &(mag) &0.416 &0.488 &0.415 &0.533 &0.457 \\
eE(B--V)  &(mag) &0.0100 &0.0080 &0.0160 &0.0180 &0.0120 \\
 phot\_bp\_rp\_excess\_factor  &(mag) &1.2363 &1.2715 &1.2109 &1.2444 &1.2291 \\
 $C^*$  & &0.0183 &0.0237 &0.0054 &0.0272 &0.0197 \\
 (BP-RP)$_0$  &(mag) &0.5634 &0.6450 &0.4397 &0.4590 &0.4592 \\
 M$_{\rm G0}$  &(mag) &0.6253 &0.6886 &0.7261 &0.6369 &0.8232 \\
 (BP-RP)$_{0,SFD}$  &(mag) &0.4540 &0.5394 &0.3280 &0.3517 &0.3494 \\
 M$_{\rm G0,SFD}$  &(mag) &0.4035 &0.4742 &0.5007 &0.4207 &0.6015 \\
 e(BP-RP)$_0$  &(mag) &0.0175 &0.0217 &0.0170 &0.0245 &0.0151 \\
 phot\_g\_mean\_mag  &(mag) &16.0204 &16.3375 &16.0999 &16.2718 &16.2846 \\
 phot\_g\_mean\_flux\_over\_error  & &67.29 &90.72 &147.96 &150.94 &322.98 \\
 phot\_bp\_mean\_mag  &(mag) &16.4255 &16.8955 &16.445 &16.6643 &16.6395 \\
 phot\_bp\_mean\_flux\_over\_error  & &18.07 &32.56 &41.38 &40.99 &85.1 \\
 phot\_rp\_mean\_mag  &(mag) &15.4123 &15.5903 &15.5742 &15.6604 &15.7231 \\
 phot\_rp\_mean\_flux\_over\_error  & &29.19 &32.13 &66.64 &60.84 &118.01 \\
 int\_average\_g  &(mag) &15.9899 &16.2077 &16.103 &16.2834 &16.2963 \\
 int\_average\_g\_error  &(mag) &0.0087 &0.0068 &0.002 &0.0086 &0.0035 \\
 int\_average\_bp  &(mag) &16.3569 &16.6623 &16.4217 &16.6724 &16.6527 \\
 int\_average\_bp\_error  &(mag) &0.0112 &0.0139 &0.0054 &0.0142 &0.0079 \\
 int\_average\_rp  &(mag) &15.3865 &15.5365 &15.5672 &15.6679 &15.7322 \\
 int\_average\_rp\_error  &(mag) &0.0090 &0.0147 &0.0024 &0.0086 &0.0046 \\
 zp\_mag\_g  &(mag) &16.0318 &16.2436 &16.1112 &16.2905 &16.2972 \\
 zp\_mag\_bp  &(mag) &16.4225 &16.7174 &16.4325 &16.6824 &16.6541 \\
 zp\_mag\_rp  &(mag) &15.4118 &15.5578 &15.5710 &15.6718 &15.7320 \\
 $F_{\rm G}$  &(mag) &16.0388 &16.2253 &16.1055 &16.2905 &16.2972 \\
 $e F_{\rm G}$  &(mag) &0.0004 &0.0006 &0.0004 &0.0005 &0.0004 \\
 $F_{\rm BP}$  &(mag) &16.4389 &16.6908 &16.4379 &16.6776 &16.6541 \\
 $e F_{\rm BP}$  &(mag) &0.0018 &0.0022 &0.0017 &0.0021 &0.0020 \\
 $F_{\rm RP}$  &(mag) &15.4228 &15.5450 &15.5715 &15.6718 &15.7341 \\
 $e F_{\rm RP}$  &(mag) &0.0013 &0.0015 &0.0014 &0.0016 &0.0016 \\
 \enddata 
 \tablecomments{$F_{\rm G}$, $F_{\rm BP}$, and $F_{\rm RP}$ are the Fourier zero point brightness values from our own analysis.}
\end{deluxetable*}


\begin{acknowledgments}
CsK thanks the organizers, mentors, and participants of the MWGaiaDN Winter School in Leiden, 2024, for the fruitful discussions on \textit{Gaia} extinction calculations. CsK was also supported by the EKÖP-24 University Excellence Scholarship Program of the Ministry for Culture and Innovation, financed by the the National Research, Development and Innovation Fund. This research was supported by the `SeismoLab' KKP-137523 \'Elvonal grant of the Hungarian Research, Development and Innovation Office (NKFIH), and by the LP2025-14/2025 Lendület grant of the Hungarian Academy of Sciences. This work has made use of data from the European Space Agency (ESA) mission {\it Gaia} (\url{https://www.cosmos.esa.int/gaia}), processed by the {\it Gaia} Data Processing and Analysis Consortium (DPAC, \url{https://www.cosmos.esa.int/web/gaia/dpac/consortium}). This research was supported by the International Space Science Institute (ISSI) in Bern/Beijing through ISSI/ISSI-BJ International Team project ID \#24-603 - “EXPANDING Universe” (EXploiting Precision AstroNomical Distance INdicators in the Gaia Universe). This research makes use of public auxiliary data provided by ESA/Gaia/DPAC/CU5 and prepared by Carine Babusiaux. This research made use of NASA’s Astrophysics Data System Bibliographic Services, as well as of the SIMBAD and VizieR databases operated at CDS, Strasbourg, France.
\end{acknowledgments}

\facilities{Gaia (ESA,\citealt{GaiaCollaboration2016}}

\software{astropy \citep{astropy:2013, astropy:2018, astropy:2022}; seismolab \citep{Bodi2024}; dustmaps \citep{dustmaps-2018}; gaiadr3\_bcg
}

\bibliography{references}{}
\bibliographystyle{aasjournalv7}

\end{document}